\pdfoutput=1
\documentclass[nofootinbib]{revtex4}
\usepackage{amsmath,amssymb,graphicx}
\usepackage{epsf}
\usepackage{epstopdf}
\usepackage {amssymb}
\usepackage{color}
\newcommand{\nc}{\newcommand}
\nc{\ba}{\begin{eqnarray}}
\nc{\ea}{\end{eqnarray}}
\newcommand\be{\begin{equation}}
\newcommand\ee{\end{equation}}

\definecolor{darkgreen}{rgb}{0,0.5423,0.1}
\definecolor{darkblue}{rgb}{0,0.3,0.943}
\definecolor{darkred}{rgb}{0.6,0,0}

\nc{\e}{{\bf{e}}}
\nc{\kk}{{\bf{k}}}
\nc{\pp}{{\bf{p}}}

\begin{document}
\title{CMB statistical anisotropies of classical and quantum origins}

\author{Xingang Chen$^1$}
\email{Xingang.Chen-AT-utdallas.edu}

\author{Razieh Emami$^2$}
\email{emami-AT-ipm.ir}

\author{Hassan Firouzjahi$^3$}
\email{firouz-AT-ipm.ir}

\author{Yi Wang$^4$}
\email{yw366-AT-cam.ac.uk}

\affiliation{$^1$Department of Physics, The University of Texas at Dallas, Richardson, TX 75083, USA}

\affiliation{$^2$School of Physics, Institute for Research in
Fundamental Sciences (IPM)
P.~O.~Box 19395-5531,
Tehran, Iran}

\affiliation{$^3$School of Astronomy, Institute for Research in
Fundamental Sciences (IPM)
P.~O.~Box 19395-5531,
Tehran, Iran}

\affiliation{$^4$Centre for Theoretical Cosmology, DAMTP, University of Cambridge, Cambridge CB3 0WA, UK}

\begin{abstract}
We examine the impact of different anisotropic relics on inflation, in particular the predictions on the density perturbations. These relics can be the source of the large scale anomalies in the cosmic microwave background. There are two different types of background relics, one from the matter sector and the other purely from the metric. Although the angular-dependence of the statistical anisotropy in both cases are degenerate, the scale-dependence are observationally distinctive. In addition, we demonstrate that non-Bunch-Davies vacuum states can extend the statistical anisotropy to much shorter scales, and leave a scale-dependence that is insensitive to the different backgrounds but sensitive to the initial quantum state.

\end{abstract}

\maketitle


\section{Introduction}

Several anomalies in the largest scales of the CMB \cite{Bennett:2010jb,Ade:2013nlj} have been interesting sources of inspiration for constructing models of  early universe beyond the Standard Model of cosmology. For example, it has been found that there are certain scale-dependent statistical anisotropies in these scales \cite{Tegmark:2003ve,de OliveiraCosta:2003pu,Schwarz:2004gk,Land:2005ad,Copi:2006tu}. Before inflation, the universe is supposed to be inhomogeneous and anisotropic. Without active supporting sources, these initial relics at the beginning of inflation are wiped out by inflation very quickly. However, if the number of e-folds of inflation is minimal, namely not much more than that is required to solve the flatness and the horizon  problems of Big Bang, these initial relics can leave their imprints in the statistics of the largest scales of the density perturbations.

These statistical anisotropies are naturally scale-dependent. The details of scale-dependence crucially rely on the background source of the anisotropy and the initial quantum states. Such relations between the early universe models and observables provide a special window to the physics of the early universe. To properly understand these relations, systematic classification of different relic models and their predictions are necessary. 
This is particularly important because the analyses of large
scale anomalies are often limited by the cosmic variance. By classifying the relic models, their predictions
are classified into packages.
These predictions include the scale dependence and angular dependence of the anomalies, together with other possible predictions on such as spatial curvature and non-Gaussianities. When comparing with data, the package of predictions provide theoretical templates which may provide a unified explanation for several anomalies. Some related new anomalies
may be predicted and verified, substantially increasing the statistical significance.
In addition, systematic studies of different anomalies in model-building can tell us not only why they are present, but also which fundamental physics we are able to probe.

With these motivations in mind, we note that there are two classes of models with initial anisotropics relics. The source of the initial anisotropy can either be matter fields, or solely from the gravitational sector\footnote{There is also a large class of models where the anisotropy has an active source. For example, inflation supported by an attractor vector field \cite{Watanabe:2009ct, Emami:2010rm}, see also  \cite{Chen:2014eua, Ohashi:2013qba, Thorsrud:2013kya} and the references therein, or bifurcation of inflationary trajectory \cite{Li:2009sp, Afshordi:2010wn, Wang:2013zz}. Alternatively, the anisotropy may not be efficiently diluted when the inflationary dynamics is modified \cite{Endlich:2012pz, Bartolo:2013msa, Akhshik:2014zz}. To distinguish, we do not call them the {\em relics} models.}.
The main goal of this paper is to compare the predictions of these two classes of models.
For the first type of models, an example of relic vector field has been studied analytically and numerically in Ref.~\cite{Chen:2013eaa,Chen:2013tna}. This model gives a specific prediction on the form and scale-dependence of the statistical anisotropy of the CMB. The dependence of the prediction on the initial quantum fluctuation state is also studied. For the second type, a Bianchi-type inflationary background model has been studied in \cite{Gumrukcuoglu:2007bx,Pereira:2007yy}. The density perturbations in this study was done only numerically. To properly compare them with the first type of models and to make the prediction more relevant to the data analyses, we use the same perturbative method as in \cite{Chen:2013eaa,Chen:2013tna} to solve these models analytically. We examine the angular and scale-dependence of the statistical anisotropy in these two types of models. In addition, we study the effects of the initial quantum state on these predictions following \cite{Chen:2013eaa,Chen:2013tna}, and emphasize how the resulting distinctive scale-dependence can be used as a probe of non-Bunch-Davies (non-BD) vacuum. For earlier works considering anisotropies generated from initial anisotropies in metric with non-BD vacuum see \cite{Dey:2013tfa, Dey:2011mj, Dey:2012qp}.
The imprint of non-BD initial condition in models of anisotropic inflation \cite{Watanabe:2009ct} was
also studied in \cite{Emami:2014tpa}.

\section{Background evolution}
\label{model}

We start with the minimal model of inflation based on a scalar field minimally coupled to gravity
\ba
\label{action} S= \int
d^4 x  \sqrt{-g} \left [ \frac{M_P^2}{2} R - \frac{1}{2} \partial_\mu \phi
\,  \partial^\mu \phi - V(\phi) \right] \, ,
\ea
in which $M_P$ is the reduced Planck mass.

Before inflation reaches its attractor isotropic FRW phase, the expansion rates along different spatial directions may be different. The difference can be modeled by the type I Bianchi Universe, with the metric
\ba
\label{bian0}
ds^2 = - dt^2 + a^2 d x^2 + b^2(d y^2 +d z^2) \, .
\ea
Note that to simplify the analysis, we have assumed that there is a remnant two-dimensional symmetry in
$y-z$ plane. Later on we consider the most general case in which all three directions are anisotropic.

Considering the following ansatz for the scale factors $a$ and $b$, $a \equiv e^{\alpha(t)}$ and $b \equiv e^{\alpha(t)+3\sigma(t)}$, the metric (\ref{bian0}) becomes
\ba
\label{bian01}
ds^2 &=& - dt^2 + e^{2\alpha(t)}\left(d x^2
+e^{6\sigma(t)}(d y^2 +d z^2) \right) \, .
\ea
With the metric in this form, the background field equations are
\ba
\label{back-rho-eq}
\ddot\phi+3\left(\dot \alpha + 2 \dot \sigma \right)\dot \phi+ V_\phi &=&0  \\
\label{Ein1-eq}
3 M_P^2 \left(\dot \alpha^2+ 4\dot \alpha \dot \sigma + 3\dot \sigma^2 \right) &=& \frac{1}{2}\dot
\phi^2+V(\phi) \\
\label{Ein2-eq}
M_P^2 \left( \ddot \alpha + 3\dot \alpha \left( \dot \alpha + 2 \dot \sigma \right) \right) &=& V(\phi) \\
\label{anisotropy-eq}
\ddot \sigma +3\dot \sigma \left( \dot \alpha + 2 \dot \sigma \right)&=& 0\, ,
\ea
in which a dot indicates derivative with respect to $t$.

One can integrate the above equations and to leading order in slow-roll expansion obtain (the details can be found in appendix \ref{slow-roll} )
\ba
\label{app.a}
a & \simeq & H_{0}^{-1} \left( -\eta \right)^{-1} \\
\label{app.b}
b & \simeq & H_{0}^{-1} \left( -\eta \right)^{-1}\left( 1 + \left(\frac{\dot{\sigma_{0}}}{H_{0}}\right) \left( \mathcal{H}_{0} \eta \right)^3 \right) ~,
\ea
in which the subscript $0$ represents the values of the corresponding quantities at the start of inflation
$\eta =\eta_0$, $H= \dot \alpha$ is the leading order Hubble expansion rate and $\mathcal{H}  \equiv a H$.

\section{Perturbations}

Here we study perturbations in this model. The perturbation in this model is solved numerically in \cite{Gumrukcuoglu:2007bx}. However, in order to compare the results with a different model presented in \cite{Chen:2013tna, Chen:2013eaa}, here we solve the model analytically as in \cite{Chen:2013tna, Chen:2013eaa}. In principle one should take into account the perturbations in both of the matter and metric sectors. This can be achieved by integrating out the non-dynamical degrees of freedom. We leave the details of this analysis to appendix \ref{metric pert}. However, due to the slow-roll approximation, it turns out that the additional terms from integrating out the metric degrees of freedom are sub-leading compared to the typical terms coming from the matter sector and in order to read off the leading corrections we can neglect them all together \cite{Emami:2013bk}.

Neglecting the metric perturbations,  the second order action for $\delta \phi$ is then well approximated by (see Appendix \ref{quad-action} for the total form of the quadratic action)
\ba
L_{\phi \phi} \simeq \frac{b^2}{2}\mid \delta \phi_{k} '\mid^2
- \left[ \frac{b^2}{2} k_x^2 + \frac{a^2}{2} (k_y^2 + k_z^2) \right]
\mid \delta \phi_{k}\mid^2
- \frac{a^2 b^2}{2} V_{,\phi\phi}\mid \delta \phi_{k}\mid^2 ~,
\ea
where the last term is also slow-roll suppressed and can be neglected.
Throughout the paper we use the
prime to indicate the derivative with respect to the conformal time defined in terms of the scale factor $a(t)$, $d \eta = dt/a(t)$.
The equation of motion for  $\delta \phi$ in Fourier space is
\ba
\label{KG1}
\delta \phi_{k}'' + 2 \frac{b'}{b}\delta \phi_{k}' + \left( k_{x}^2 + \frac{a^2}{b^2}\left(k_{y}^2 + k_{z}^2 \right)\right)\delta \phi_{k} =0 \, .
\ea
We can expand $\delta \phi $ in terms of the usual creation and annihilation operators as
\ba
\label{a-adag}
\delta \phi = \int \frac{d^3k}{(2\pi)^3}\bigg{[}u_{k}a_{\kk} + u_{k}^{*}a_{-\kk}^{\dag}\bigg{]}e^{i\mathbf{k.x} } \equiv  \int \frac{d^3k}{(2\pi)^3} \delta \phi_{\kk} e^{i \kk \cdot {\bf x} } \, .
\ea
By using Eqs. (\ref{app.a}-\ref{app.b}) and the above expansion, the perturbed scalar field equation (\ref{KG1}) is written as
\ba
\label{mode function}
u_{k}'' -\frac{2}{\eta}\left(1 - 3 \left(\frac{\dot{\sigma_{0}}}{H_{0}}\right)\left(\mathcal{H}_{0}\eta\right)^3  \right)u_{k}' + \left( k_{x}^2 + \left( 1 - 2 \left(\frac{\dot{\sigma_{0}}}{H_{0}}\right) \left(\mathcal{H}_{0}\eta\right)^3 \right)\left(k_{y}^2 + k_{z}^2 \right)\right)u_{k} =0 ~.
\ea

In this paper we are interested in small anisotropies so we can solve the above equation perturbatively. Since the effect of Bianchi anisotropy has been parameterized by $\frac{\sigma_{0}'}{\mathcal{H}_{0}}\left(\mathcal{H}_{0}\eta\right)^3$, we would expect that all modes, either near the horizon or well inside the horizon, are affected by the anisotropy of this order. However, as shown in \cite{Chen:2013tna}, in order to see this explicitly
a proper change of variables in (\ref{mode function}) is necessary. In the following, first we solve equation \eqref{mode function} using the original  variable $u_k$. As we will see, the expansion breaks down for modes deep inside the horizon. We improve our expansion scheme by changing to a new variable and present an expansion which is suitable for both near horizon and UV modes.

\subsection{Near Horizon expansion}
Now we would like to solve the equation of motion for perturbations. Following \cite{Chen:2013tna, Chen:2013eaa}, we can expand $u_{k}$ as
\ba
\label{uk}
u_{k} = \mathcal{C}_{+}u_{k(0)} + u_{k(1)} ~,
\ea
in which the zeroth order isotropic wave function is given by
\ba
 u_{k(0)} = \frac{H_{0}}{\sqrt{2k^3}}\left( 1+ ik\eta \right)e^{-ik\eta} ~.
\ea
One can interpret $\mathcal{C}_{+}$ as the correction in wave function normalization and $u_{k(1)}$ as the corrections in the profile of wave function in the presence of anisotropy.

The next order $u_{k(1)}$ can be solved perturbatively from the following equation,
\ba
u_{k(1)}'' -\frac{2}{\eta}u_{k(1)}' + k^2 u_{k(1)} = - \frac{6}{\eta}\left(\frac{\dot{\sigma_{0}}}{H_{0}}\right)
\left(\mathcal{H}_{0}\eta\right)^3u_{k(0)}'+ 2\left(\frac{\dot{\sigma_{0}}}{H_{0}}\right) \left(\mathcal{H}_{0}\eta\right)^3 \left(k_{y}^2 + k_{z}^2 \right)u_{k(0)} ~.
\ea
Using the ansatz
\ba
u_{k(1)} = \frac{\dot{\sigma_{0}}}{\sqrt{2k^3}}\mathcal{H}_{0}^3 \sum_{n=3}^{5} \alpha_{n}\eta^{n} e^{-ik\eta}
\ea
we get
\ba
\alpha_{3} &=& - \frac{1}{4k^2}\left( 4k_{x}^2 + k_{y}^2 + k_{z}^2 \right) \\
\alpha_{4} &=& - \frac{i}{4k}\left( 4k_{x}^2 + k_{y}^2 + k_{z}^2 \right) \\
\alpha_{5} &=& - \frac{1}{4}\left( k_{y}^2 + k_{z}^2 \right)
\label{alpha5value}
\ea
We see that for $k\eta > \left(\frac{\dot{\sigma_{0}}}{H_{0}}\right)^{-1}\left( \mathcal{H}_{0} \eta\right)^{-3}$ the above expansion breaks down, as we discussed.
We will come back to this point soon.

We determine $\mathcal{C}_{+}$ by using the following normalization condition
\ba
[\delta \phi_{\mathbf{q}}, \delta \pi_{\mathbf{p}}] = i (2\pi)^3 \delta^3(\mathbf{p}+ \mathbf{q}) \, ,
\ea
where $\delta \pi_{\mathbf{p}}$ is the momentum conjugate associate with
$\delta \phi_{\mathbf{p}}$,  $\delta \pi_{\mathbf{p}} = b^2 \delta \phi'_{\mathbf{p}}$. The above condition leads to the following equation
\ba
\frac{1}{H_{0}^2\eta^2}\left( 1 + 2 \left(\frac{\dot{\sigma_{0}}}{H_{0}}\right)\left( \mathcal{H}_{0} \eta\right)^3 \right) \left( u_{q} u_{q}^{'*} - u_{q}^{*} u_{q}^{'} \right) = i \, .
\ea
Since $\eta \to 0 $, we just keep the leading constant term. It turns out that only $\alpha_{3}$ plays role while the other higher terms are exponentially suppressed.
We get
\ba
\label{c plus0}
|\mathcal{C}_{+}|^2 =  1 + \frac{3\mathcal{H}_{0}^2}{4k^3}\sigma_{0}' \left(1 + 3 \cos^2{\Theta} \right)  \, ,
\ea
in which the amplitude of momentum $k$ and the angle $\Theta $ are defined as
\ba
k^2 \equiv k_x^2 + k_y^2 + k_z^2 ~, ~~~
\cos\Theta \equiv k_x/k ~.
\label{k_definition}
\ea

\subsection{UV safe expansion}

One might have some doubts in the above expansion scheme because
it breaks down for short wavelength modes $k\eta > \left(\frac{\dot{\sigma_{0}}}{H_{0}}\right)^{-1}\left( \mathcal{H}_{0} \eta\right)^{-3}$ due to the last term (\ref{alpha5value}).
Physically we do not expect this to happen. This problem is especially important if we would like to study the effect of anisotropic relics on the short wavelength modes. So to demonstrate explicitly the validity of our method, we have to elaborate the expansion scheme. It turns out that this can be fixed by properly choosing the variable used in the perturbative method. The expansion will be perturbative for all modes if we choose to perturbatively expand the exponent in the variable $u_k$ \cite{Chen:2013tna}.
Defining
\ba
\label{psi u}
\psi_{\mathbf{k}}(\eta)\equiv \log{\left(u_{\mathbf{k}}(\eta)\right)} \, ,
\ea
we expand $\psi_{\mathbf{k}}(\eta)$ in orders of $\sigma_{0}'$
\ba
\psi_{\mathbf{k}}(\eta) = \psi_{\mathbf{k}(0)}(\eta)+ \psi_{\mathbf{k}(1)}(\eta)+ ... ~.
\ea
One can then solve this perturbatively (see Appendix \ref{psi solution}) and get
\begin{align}
\label{psi zeroth}
\psi_{\mathbf{k}(0)}(\eta) &= \log{\left(u_{\mathbf{k}(0)}(\eta)\right)}
\\
\psi_{\mathbf{k}(1)}(\eta) &= \frac{3i}{8k^3} \mathcal{H}_{0}^2\sigma_{0}' \left(1 + 3 \cos^2{\Theta}\right) +\frac{ \mathcal{H}_{0}^2\sigma_{0}' }{1+ik\eta}\sum_{n=3}^{5} \alpha_{n}\eta^{n}\nonumber\\
&\equiv \frac{\mathcal{H}_{0}^2 \sigma_{0}'}{1+ik\eta}\sum_{n=0}^{5} \beta_{n}\eta^{n} \, ,
\end{align}
where $\beta_{n}$ are given by
\ba
\beta_{0} &=& \frac{3i}{8k^3}\left(1 + 3 \cos^2{\Theta}\right) \\
\beta_{1} &=& -\frac{3}{8k^2}\left(1 + 3 \cos^2{\Theta}\right) \\
\beta_{2} &=& 0 \\
\beta_{m} &=& \alpha_{m} ~~,~~ (m=3,4,5) \, .
\ea
For UV modes, $\psi_{k(0)} \sim -ik\eta$ and
$\psi_{k(1)} \sim (\sigma'_0 \mathcal{H}_0^2 \eta^3) k\eta$.
So the anisotropic corrections remain small for all modes.

At late time, the conserved curvature perturbation approaches to the attractor single field expression, $\zeta \approx - H_{0} \left(\frac{\delta \phi}{\dot \phi_{0}}\right)$, in the gauge used here. Therefore we can use this time-delay formula to compute the power spectrum by evaluating the variables at their attractor values. The statistical anisotropy in the finite result shows up through the coefficient ${\cal C}_+$ we just computed. We thus have
\ba
\label{power-anis}
\left( \frac{k^3}{2 \pi^2} \right)  \big \langle \zeta^2   \big \rangle = P_{\zeta0}\left( 1 + \frac{3\mathcal{H}_{0}^2}{4k^3}\sigma_{0}' \left(1 + 3 \cos^2{\Theta} \right) \right) \, ,
\ea
where the isotropic power spectrum is defined via  $P_{\zeta0} \equiv \frac{H_{0}^4}{\left(2\pi\dot{\phi_{0}} \right)^2}$.

Now we can compare this result with that in the model of relic vector field \cite{Chen:2013tna,Chen:2013eaa}. In both models, the anisotropy is axial symmetric, so as expected they have the same angular-dependence. But due to the different sources, the anisotropies in these two types of models have different scale-dependence. In the relic vector case the anisotropy decays towards smaller scales as $\sim 1/k^4$. However, here in the Bianchi type cases in which  anisotropy is generated from anisotropic scale factors, it decays as $\sim 1/k^3$. These two different behavior are related to the different decay speeds of the background relics in the models.


\section{An non-BD example: Gaussian state}

For inflation with minimal number of e-folds, the initial state of quantum fluctuations also do not have to be in their attractor vacuum states. It is therefore a sensible question to consider the effects of the non-BD states, and see how the initial quantum states of the universe leave their imprints in the statistical anisotropy of the CMB \cite{Chen:2013tna, Emami:2014tpa}. Conversely any distinctive predictions can then be used as a probe of the initial quantum state of the Universe.
In the following, we use a specific example for the non-BD vacuum, namely the Gaussian state [\citenum{Polarski:1995jg},\citenum{Chen:2013tna}]\footnote{There are other proposals and methods to model \cite{Barrow:1997sy, Barrow:1998ih} and probe the initial non-BD states \cite{Brandenberger:2000wr,Easther:2001fi,Chen:2006nt,Holman:2007na,Meerburg:2009ys,Chen:2009bc,Agullo:2010ws,Ganc:2011dy,Chialva:2011hc,Berezhiani:2014kga}.}. As we will see, there are two types of scale-dependence. One has an oscillatory behavior while the other is not oscillatory.
To start, let us write down the quadratic Hamiltonian for the quantum fluctuations of the inflaton field, $\delta \phi$, in a canonical form
\ba
v_\mathbf{k} &=& b \delta \phi_\mathbf{k} \\
\pi_\mathbf{k} &=& v_\mathbf{k}' - \frac{b'}{b}v_\mathbf{k}
\ea
The Hamiltonian is
\ba
H_2 = \int \frac{d^3 \mathbf{k}}{(2\pi)^3} \frac{1}{2} \left[ \left(\pi_\mathbf{k} \pi^{*}_\mathbf{k}\right) + \frac{\left(k_y^2 + k_z^2 + k_x^2 \left( 1 + \sigma_{0}' \mathcal{H}_{0}^2 \eta^3\right)^2\right)}{\left( 1 + \sigma_{0}' \mathcal{H}_{0}^2 \eta^3\right)^2} \left(v_\mathbf{k} v^{*}_\mathbf{k} \right)+ \frac{\left(-1 + 2\sigma_{0}'\mathcal{H}_{0}^2\eta^3\right)}{\eta \left( 1 + \sigma_{0}' \mathcal{H}_{0}^2 \eta^3 \right)}\left(\pi_\mathbf{k} v^{*}_\mathbf{k} + \pi^{*}_\mathbf{k} v_\mathbf{k}  \right) \right] \, .
\ea
Using the Schrodinger picture to quantize the fields as
\ba
\label{wavefunction}
v_\mathbf{k} &=& f_\mathbf{k}(\eta) a_\mathbf{k}(\eta_0) + f^{*}_\mathbf{k}(\eta) a^{\dag}_{-\mathbf{k}}(\eta_0), \nonumber\\
\pi_\mathbf{k} &=& -i \left(g_\mathbf{k}(\eta) a_\mathbf{k}(\eta_0) - g^{*}_\mathbf{k}(\eta) a^{\dag}_{-\mathbf{k}}(\eta_0)\right),
\ea
where the creation and annihilation operators satisfy the usual commutation relations and
\ba
f_\mathbf{k}(\eta) = u_{0\mathbf{k}}b(\eta)\left(C_{+0}e^{\psi_{1\mathbf{k}}} + C_{-0}e^{\psi^{*}_{1\mathbf{k}}} \right)
\ea
is proportional to mode function with $C_{+0}$ and $C_{-0}$ being initial  constants
while
\ba
g_{\mathbf{k}}(\eta) = i \left( f'_{\mathbf{k}} - \mathcal{H}_{b}f_{\mathbf{k}} \right) \, .
\ea
Now we can define the Gaussian state at $\eta_0$ as,
\ba
a_{\mathbf{k}}(\eta_0) \mid 0, \eta_0 \rangle =0 ~.
\ea
Through this condition, the initial quantum state acquires an anisotropic component due to the anisotropic background.
By using Eq. (\ref{wavefunction}), we get
\ba
\label{condition1}
(g_{\mathbf{k}} - i \widehat{k} v_{\mathbf{k}})\mid_{\eta_{0}} =0 \, ,
\ea
where we have defined,
\ba
\label{hatk}
\widehat{k}^2 \equiv \frac{\left(k_y^2 + k_z^2 + k_x^2 \left( 1 + \sigma_{0}' \mathcal{H}_{0}^2 \eta^3\right)^2\right)}{\left( 1 + \sigma_{0}' \mathcal{H}_{0}^2 \eta^3\right)^2}
\ea
On the other hand, we may also use the normalization condition for $C_{+0}$ and $C_{-0}$ as
\ba
\label{condition2}
\mid C_{+0}\mid ^2 - \mid C_{-0}\mid ^2 =1.
\ea
Now by using Eqs. (\ref{condition1}) and (\ref{condition2}), the power spectrum is proportional to
\ba
\mid C_{+0} + C_{-0} \mid ^2 e^{2\psi_{\mathbf{k}(1)}\mid_{\eta\rightarrow 0}} = 1 + \frac{1}{2k^2\eta_0^2} + \sigma_{0}' \mathcal{H}_{0}^2 \left( - \frac{\eta_0}{2k^2} - \frac{5\eta_0}{2k^2}\cos^2{\Theta} \right) + {\rm oscillation ~~ terms},
\label{non-BD_results}
\ea
where we have the following expression for the oscillation terms
\ba
{\rm oscillation~~ terms} &=& \left[ -\frac{1}{2k^2\eta_0^2} + \sigma_{0}' \mathcal{H}_{0}^2 \left( \frac{3}{4k^4\eta_0} + \frac{\eta_0}{2k^2} + \frac{\eta_{0}^3}{2} + \cos^2{\Theta}\left( \frac{9}{4k^4 \eta_{0}} + \frac{5\eta_{0}}{2k^2}-\frac{\eta_{0}^3}{2}\right) \right) \right]\cos{2k\eta_{0}} \nonumber\\
&+&\left[ - \frac{1}{k\eta_0} + \sigma_{0}' \mathcal{H}_{0}^2 \left( \frac{-3 + 2k^4\eta_{0}^4}{8k^5 \eta_{0}^2} + \cos^2{\Theta} \left( \frac{ -9 + 22 k^4 \eta_{0}^4}{ 8 k^5 \eta_{0}^2} \right)
\right) \right]\sin{2k\eta_{0}}
\ea
We are mostly interested in the non-oscillatory anisotropic terms because such terms are sensitive probes of the initial quantum states \cite{Chen:2013tna, Chen:2013eaa}.\footnote{If $\eta_0$ is at the beginning of inflation, the frequency of the oscillatory components is high and approaches the ultimate resolution of CMB.} Interestingly, while in the BD cases, the scale-dependence of the statistical anisotropies are different for the relic vector field model  ($\sim 1/k^4$) \cite{Chen:2013tna, Chen:2013eaa} and the Bianchi model ($\sim 1/k^3$) as we obtained in previous Section, the effect of the non-BD Gaussian state on both models are the same. Such a state generically extends the anisotropy to much smaller scales and the scale-dependence
for both cases are  $\sim 1/k^2$. This can also be readily understood. In the BD case, the background evolution plays the dominant role in the final results. However, in the non-BD case the initial quantum states play more important roles enhancing the anisotropy of the shorter wave-length modes.
We have given just one example of non-BD state here. It is plausible that the enhancement caused by other non-BD states can have different scale-dependence.


\section{Generalization: full anisotropy in all 3 spatial directions}
\label{total Anisotrop}

In the previous sections we have reduced the three-dimensional spatial translational symmetry to the
two-dimensional translational symmetry. In the rest of the paper, we generalize these results to the maximally anisotropic case in which all three scale factors are different.
We expect the scale-dependence of the anisotropic power spectrum
to be the same as we studied above. However, we expect the angular-dependence to be different.

\subsection{Background }

In this case the background metric is given by
\ba
\label{biantotal}
 ds^2 = - dt^2 + a^2 d x^2 + b^2d y^2 + c^2d z^2 \, .
\ea
Considering the following anasatz for the scale factors $a$, $b$ and $c$,
\ba
\label{abc}
a(t)&=& e^{\alpha(t)} \\
b(t)&=& e^{\alpha(t)+ 3\sigma(t)}\\
c(t)&=& e^{\alpha(t)+ 3\delta(t)} \, ,
\ea
the background equations of motion is
\ba
\label{back-rho-eq-to}
\ddot\phi+3\left(\dot \alpha + \dot \sigma + \dot \delta \right)\dot \phi+ V_\phi &=&0  \\
\label{Ein1-eq-to}
3 M_P^2 \left(\dot \alpha^2+ 2\dot \alpha \left(\dot \sigma + \dot \delta \right) + 3\dot \sigma \dot \delta\right) &=& \frac{1}{2}\dot
\phi^2+V(\phi) \\
\label{Ein2-eq-to}
M_P^2 \left( \ddot \alpha + 3\dot \alpha \left( \dot \alpha +  \dot \sigma + \dot \delta \right) \right) &=& V(\phi) \\
\label{anisotropy-eqy-to}
\ddot \sigma +3\dot \sigma \left(\dot \alpha +  \dot \sigma + \dot \delta  \right)&=& 0\, ,
\\
\label{anisotropy-eqz-to}
\ddot \delta +3\dot \delta \left(\dot \alpha +  \dot \sigma + \dot \delta  \right)&=& 0\, .
\ea
\begin{figure}[t]
\includegraphics[width=0.6\textwidth]{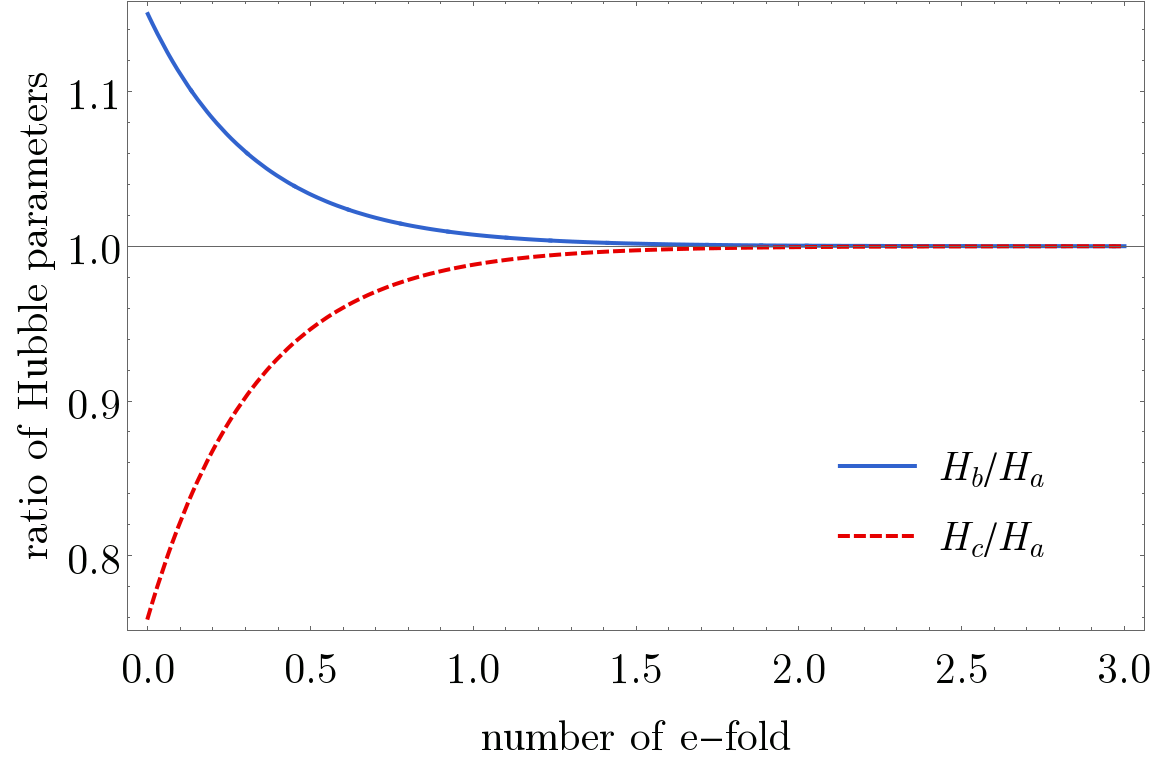}
 \caption{Here we plot the evolution of $H_{b}$ and $H_{c}$. As we expect, the attractor solution is FRW as the system approaches to it very rapidly. The parameters are chosen such that at an initial time $t_0$, $\dot \sigma_{0} = 0.05 \dot \alpha_{0}$ and $\dot \delta_{0} = -0.08 \dot \alpha_{0}$.}
 \vspace{0.5cm}
\label{Fattractor}
\end{figure}
Although the above equations seem to be complicated, they can be simplified by using the slow-roll approximation. The situation is similar to the previous case where both of $\dot \sigma$ and $\dot \delta$ decay like $a^{-3}$ and our system approaches to its attractor   FRW phase. We present the attractor solutions in Fig.~\ref{Fattractor}.

As in our previous case, we can integrate the above equations and find the following approximate solutions
\ba
\label{app.ato}
a & \simeq & H_{0}^{-1} \left( -\eta \right)^{-1 } \\
\label{app.bto}
b & \simeq & H_{0}^{-1} \left( -\eta \right)^{-1}\left( 1 + \left(\frac{\dot{\sigma_{0}}}{H_{0}}\right) \left( \mathcal{H}_{0} \eta \right)^3 \right) \\
\label{app.cto}
c & \simeq & H_{0}^{-1} \left( -\eta \right)^{-1}\left( 1 + \left(\frac{\dot{\delta_{0}}}{H_{0}}\right) \left( \mathcal{H}_{0} \eta \right)^3 \right) \, .
\ea


\subsection{Perturbations of the fully anisotropic background}

Now we consider the perturbations of our full anisotropic background. As we have justified before, we can safely neglect the metric perturbations and only consider the inflaton fluctuations. Then the second order action is
\ba
L_{\phi \phi} = \frac{bc}{2}\mid \delta \phi_{k} '\mid^2  - \left(\frac{bc}{2}k_{x}^2 + \frac{a^2c}{2b}k_{y}^2 +\frac{a^2b}{2c}k_{z}^2 \right)\mid \delta \phi_{k}\mid^2 ~.
\ea
Again we have neglected the terms that are slow-roll suppressed. Now the equation of motion for $\delta \phi$ is
\ba
\delta \phi_{k}'' + \left(\frac{b'}{b} + \frac{c'}{c}\right)\delta \phi_{k}' + \left( k_{x}^2 + \frac{a^2}{b^2}k_{y}^2 + \frac{a^2}{c^2}k_{z}^2\right)\delta \phi_{k} =0 \, .
\ea
Expanding $\delta \phi $ in terms of usual creation and annihilation operators as in Eq. (\ref{a-adag}),  the perturbed scalar field equation becomes
\ba
\label{mode functionto}
u_{k}''  &-&\frac{1}{\eta}\left[ 2 - 3 \left(\frac{\dot{\sigma_{0}}}{H_{0}}\right)\left(\mathcal{H}_{0}\eta\right)^3  - 3 \left(\frac{\dot{\delta_{0}}}{H_{0}}\right)\left(\mathcal{H}_{0}\eta\right)^3 \right]u_{k}' \nonumber\\
&&+ \bigg{[}k_{x}^2 + \left( 1 - 2 \left(\frac{\dot{\sigma_{0}}}{H_{0}}\right) \left(\mathcal{H}_{0}\eta\right)^3 \right)k_{y}^2
+ \left( 1 - 2 \left(\frac{\dot{\delta_{0}}}{H_{0}}\right) \left(\mathcal{H}_{0}\eta\right)^3 \right)k_{z}^2 \bigg{]}u_{k} =0\, .
\ea
Parallel to what we did for the axial-symmetric case,
first we solve the above equation for the ``near to horizon modes" and then we improve our expansion  by changing the variable. Subsequently, we  present an expansion which is suitable for both of near horizon and UV modes.

\subsection{Generalized near horizon expansion}

Following our previous procedure, we can expand $u_{k}$ as in Eq.~(\ref{uk}). Then the goal is finding $C_{+}$ and $u_{k(1)}$.
Let us start with the differential equation of motion for $u_{k(1)}$,
\ba
u_{k(1)}'' -\frac{2}{\eta}u_{k(1)}' + k^2 u_{k(1)} = - \frac{3}{\eta}\left(\frac{\dot{\sigma_{0}}+\dot{\delta_{0}}}{H_{0}}\right)
\left(\mathcal{H}_{0}\eta\right)^3u_{k(0)}'+ 2\left(\frac{\dot{\sigma_{0}}}{H_{0}}k_{y}^2 + \frac{\dot{\delta_{0}}}{H_{0}}k_{z}^2 \right) \left(\mathcal{H}_{0}\eta\right)^3 u_{k(0)} ~,
\ea
from which we get
\ba
u_{k(1)} = \frac{\mathcal{H}_{0}^3}{\sqrt{2k^3}}\sum_{n=3}^{5} \Omega_{n}\eta^{n}
e^{-ik\eta} \, ,
\ea
where
\ba
\Omega_{3} &=& - \frac{1}{4k^2}\left( 2k^2 \left(\dot{\sigma_{0}}+ \dot{\delta_{0}} \right)-3\left(\dot{\sigma_{0}} k_{y}^2 + \dot{\delta_{0}}k_{z}^2 \right) \right)\\
\Omega_{4} &=& - \frac{i}{4k}\left( 2k^2 \left(\dot{\sigma_{0}}+ \dot{\delta_{0}} \right)-3\left(\dot{\sigma_{0}} k_{y}^2 + \dot{\delta_{0}}k_{z}^2 \right) \right) \\
\Omega_{5} &=& - \frac{1}{4}\left(\dot{\sigma_{0}} k_{y}^2 + \dot{\delta_{0}}k_{z}^2 \right)  \, .
\ea
As it has been discussed before, we determine $\mathcal{C}_{+}$ by using the normalization condition which leads to
\ba
\label{c plus}
|\mathcal{C}_{+}|^2 =  1 + \frac{3\mathcal{H}_{0}^2}{4k^3}\bigg{[} 2 \left(\sigma'_{0}+ \delta'_{0}\right)-3 \sin^2{\Theta} \left(\sigma'_{0} \cos^2{\Phi} + \delta'_{0} \sin^2{\Phi}\right)\bigg{]}
\ea
where we have chosen the wave number $\kk$ as
\ba
\kk = (k\cos{\Theta}\, , k \sin{\Theta} \cos{\Phi}\, , k \sin{\Theta} \sin{\Phi} ) \, .
\ea
Note that as before $ \Theta $ is the angle of $\hat \kk$ with respect to the $x$ axis while
$\Phi$ is the azimuthal angle of $\hat \kk$  in $y-z$ plane.

\subsection{Generalized UV safe expansion}

As in the previous cases, we expect that all modes, including the near-horizon and UV modes, receive the same order of anistropic corrections. So we improve the expansion scheme by changing the variable as Eq. (\ref{psi u}).
Expanding $\psi_{\mathbf{k}}(\eta)$ in orders of $\dot{\sigma}_{0}$ and  $\dot{\delta}_{0}$ leads us to the following expression for $\psi_{\mathbf{k}(1)}(\eta)$
\ba
\psi_{\mathbf{k}(1)}(\eta) &=& \frac{3i\mathcal{H}_{0}^2}{8k^3}\bigg{[} 2 \left(\sigma'_{0}+ \delta'_{0}\right)-3 \sin^2{\Theta}  \left(\sigma'_{0}\cos^2{\Phi} + \delta'_{0}\sin^2{\Phi}\right)\bigg{]}
+\frac{\mathcal{H}_{0}^2 a_{0}}{1+ik\eta}\sum_{n=3}^{5} \Omega_{n}\eta^{n}\nonumber\\
&=& \frac{\mathcal{H}_{0}^2}{1+ik\eta}\sum_{n=0}^{5} \Xi_{n}\eta^{n} \, ,
\ea
where $\Xi_{n}$ are given by
\ba
\Xi_{0} &=& \frac{3i}{8k^3}\bigg{[} 2 \left(\sigma'_{0}+ \delta'_{0}\right)-3 \sin^2{\Theta} \left(\sigma'_{0} \cos^2{\Phi} + \delta'_{0} \sin^2{\Phi}\right)\bigg{]} ~,
\\
\Xi_{1} &=& -\frac{3}{8k^2}\bigg{[} 2 \left(\sigma'_{0}+ \delta'_{0}\right)-3 \sin^2{\Theta} \left(\sigma'_{0} \cos^2{\Phi} + \delta'_{0} \sin^2{\Phi}\right)\bigg{]} ~,
\\
\Xi_{2} &=& 0 ~,
\\
\Xi_{m} &=& a_{0}\Omega_{m} ~~,~~ (m=3,4,5) ~.
\ea

Using the same time-delay formula, $\zeta \approx - H_{0} \left(\frac{\delta \phi}{\dot \phi_{0}}\right)$, and the above formulas for $\delta \phi$, we can calculate the power spectrum of curvature perturbation as,
\ba\label{generalgstar}
\left( \frac{k^3}{2 \pi^2} \right)\langle \zeta^2  \rangle = P_{\zeta0}\left( 1 + \frac{3\mathcal{H}_{0}^2}{4k^3}\bigg{[} 2 \left(\sigma'_{0}+ \delta'_{0}\right)-3 \sin^2{\Theta} \left(\sigma'_{0}  \cos^2{\Phi} + \delta'_{0} \sin^2{\Phi}\right)\bigg{]}\right) \, .
\ea
Eq. (\ref{generalgstar}) is  the main result of this section and shows the non-trivial shape of anisotropic power spectrum as a function of the angles $\Theta$ and $\Phi$. In the limit where
$\sigma'_{0} = \delta'_{0}$ and $\Phi=0$, the above result coincides with the result in Eq. (\ref{power-anis}) as expected.

\begin{figure}[t]
\includegraphics[width=\textwidth]{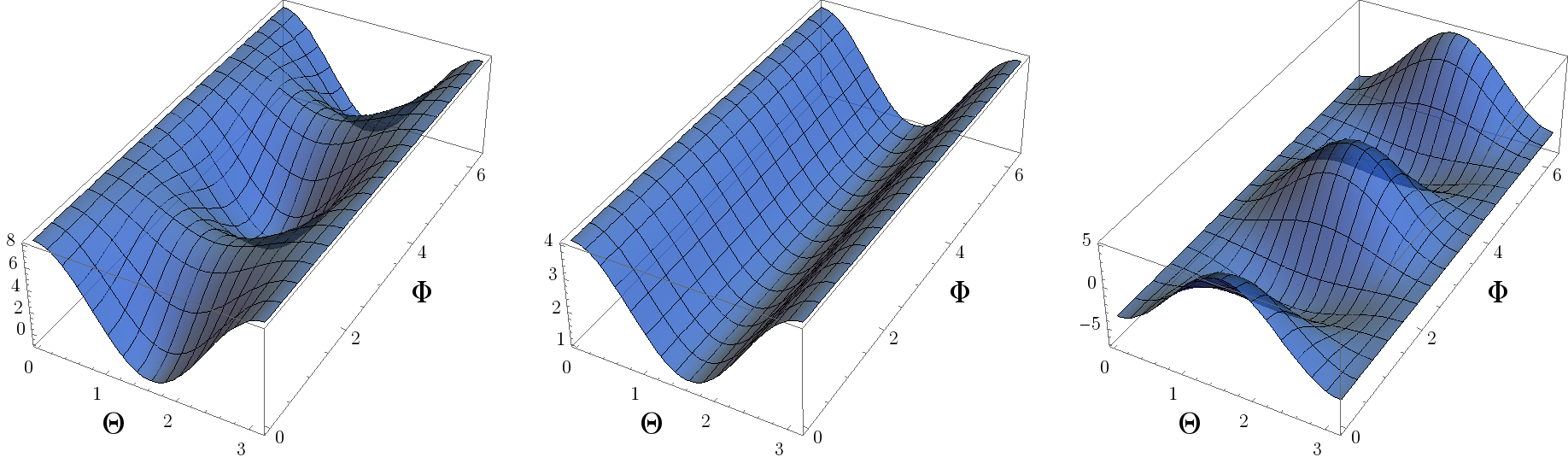}
 \caption{Here we plot $\bigg{[} 2 \left(\sigma'_{0}+ \delta'_{0}\right)-3\sin^2{\Theta} \left(\sigma'_{0} \cos^2{\Phi} + \delta'_{0} \sin^2{\Phi}\right)\bigg{]}$ for different choices of $\sigma'_{0}$ and $\delta'_{0}$. The left, middle and right panels correspond to $\sigma'_{0} = 3 \delta'_{0} $, $\sigma'_{0} = \delta'_{0} $ and $\sigma'_{0} = -3 \delta'_{0} $ respectively.}
 \vspace{0.5cm}
\label{2d angular}
\end{figure}

Since the above formula is somewhat complicated, in Fig.~\ref{2d angular} we draw  a few diagrams to illustrate this two-dimensional angular patterns. We perform this for different choices of $\sigma'_{0}$ and $\delta'_{0}$.

\section{Statistical anisotropies on the CMB}

In this section, we expand the anisotropies derived from previous sections in terms of the spherical harmonics basis. The correlation functions of the expansion coefficients are the observables on the CMB. The anisotropic corrections in Eq. (\ref{generalgstar}) are
\begin{align} \label{eq:dp}
  \Delta P_\zeta & = \frac{3\mathcal{H}_0^2}{4k^3} \left[
    2\left(\sigma_0'+\delta_0'\right) - 3  \sin^2{\Theta} \left( \sigma_0' \cos^2\Phi + \delta_0' \sin^2\Phi \right)
  \right]
  \nonumber\\ &
  = \frac{3\mathcal{H}_0^2}{4k^3} \left[
     \frac{1}{2} (1+3\cos^2\Theta) \left(\delta_0'+\sigma_0'\right)  - \frac{3}{2} (1-\cos^2\Theta)(1-2\cos^2\Phi) \left(\delta_0'-\sigma_0'\right)
  \right]~.
\end{align}
In terms of the $g_{L, M}$ parameters \cite{Pullen:2007tu} defined via
\ba
\Delta P_\zeta = \sum_{L, M} g_{L, M} Y_{LM} (\Theta, \Phi)
\ea
the  anisotropic correction can be written as
\begin{align}
  \Delta P_\zeta = P_{\zeta_0} \left[ g_{0,0}Y_{0,0} + g_{2,0}Y_{2,0} + g_{2,2}Y_{2,2} + g_{2,-2}Y_{2,-2}\right],
\end{align}
where
\begin{align}
  g_{0,0} = \frac{3\mathcal{H}_0^2}{4k^3}2\sqrt\pi \left( \delta_0'+ \sigma_0' \right)~, \quad
  g_{2,0} = \frac{3\mathcal{H}_0^2}{4k^3} \frac{2\sqrt\pi}{\sqrt 5} \left( \delta_0'+ \sigma_0' \right)~, \quad
  g_{2,2} = g_{2,-2} =  \frac{3\mathcal{H}_0^2}{4k^3} \sqrt{\frac{6\pi}{5}} \left( \delta_0'- \sigma_0' \right)~.
\end{align}
In terms of $a_{lm}$, the anisotropy can be expressed as
\begin{align} \label{eq:alm}
  a_{lm} = 4\pi (-i)^l \int \frac{d^3 k}{(2\pi)^3} ~ g_l (k) \zeta_\mathbf{k} Y^*_{lm}(\hat k)~, \qquad
  C_l = \frac{1}{2l+1} \sum_m \langle a_{lm}a^*_{lm}\rangle  \, ,
\end{align}
where $g_l(k)$ is the radiation transfer function.

Inserting the anisotropic corrections from \eqref{eq:dp},  for the anisotropic corrections  in $C_\ell$ we obtain
\begin{align}
  \Delta C_l =
  3\pi \mathcal{H}_0^2 P_{\zeta_0} \left[ \int dk ~ \frac{g_l^2(k)}{k^{4}}  \right]
  \Big\{ &  \frac{1}{2l+1}  \sum_m \int d\Omega_k ~ Y_{lm}(\mathbf{k}) Y^*_{lm}(\mathbf{k})
  \nonumber\\ &
  \times \left[
     \frac{1}{2} (1+3\cos^2\Theta) \left(\delta_0'+\sigma_0'\right)  - \frac{3}{2} (1-\cos^2\Theta)(1-2\cos^2\Phi) \left(\delta_0'-\sigma_0'\right)
  \right] \Big\} \, .
\end{align}
Note that the following integrals are $l$-independent:
\begin{align} \label{eq:aniso-integ}
  & \frac{1}{2l+1}  \sum_m \int d\Omega_k ~ Y_{lm}(\mathbf{k}) Y^*_{lm}(\mathbf{k}) \cos^2\Theta = \frac{1}{3} ~,
\nonumber\\ &
  \frac{1}{2l+1}  \sum_m \int d\Omega_k ~ Y_{lm}(\mathbf{k}) Y^*_{lm}(\mathbf{k}) \cos^2\Phi = \frac{1}{2} ~,
\nonumber\\ &
  \frac{1}{2l+1}  \sum_m \int d\Omega_k ~ Y_{lm}(\mathbf{k}) Y^*_{lm}(\mathbf{k}) \cos^2\Theta\cos^2\Phi = \frac{1}{6} ~.
\end{align}
The detail of the above calculation can be found in Appendix \ref{sec:summ-rules-spher}. As a result, the correction $\Delta C_l$ does not obtain additional $l$-dependence other than from the radiation transfer function:
\begin{align}
  \Delta C_l =
  3\pi \mathcal{H}_0^2 P_{\zeta_0} \left(\delta_0'+\sigma_0'\right) \int dk ~ \frac{g_l^2(k)}{k^{4}} ~.
\end{align}

In  the presence of statistical anisotropies, non-diagonal couplings of $\langle a_{l_1m_1}a_{l_2m_2}\rangle$ with $l_1 \neq l_2$ are turned on \cite{Ma:2011ii, Kim:2013gka, Chen:2013eaa, Watanabe:2010bu, Book:2011na}. Making use of the Gaunt's formula
\begin{align}
  & \int_0^\pi  d\Theta \int_0^{2\pi} d\Phi  ~ Y^*_{l_1m_1}Y_{l_2m_2} Y_{l_3m_3}
  \nonumber\\ =& ~
  (-1)^{m_1} \sqrt{\frac{(2l_1+1)(2l_2+1)(2l_3+1)}{4\pi}}
  \left( \begin{array}{ccc}
      l_1 & l_2 & l_3\\
      0 & 0 & 0
    \end{array} \right)
  \left( \begin{array}{ccc}
      l_1 & l_2 & l_3\\
      -m_1 & m_2 & m_3
    \end{array} \right) ~,
\end{align}
the $Y_{2,m}$ anisotropy introduces
\begin{align} \label{eq:lmlmc}
  \langle a_{l_1m_1}a^*_{l_2m_2}\rangle_{2,m}
  =
  g_{2,m}P_{\zeta_0}\frac{8\pi}{3} i^{l_2-l_1}(-1)^{m_1} \sqrt{(2l_1+1)(2l_2+1)}
  \left( \begin{array}{ccc}
      l_1 & l_2 & 2\\
      0 & 0 & 0
    \end{array} \right)
  \left( \begin{array}{ccc}
      l_1 & l_2 & 2\\
      -m_1 & m_2 & m
    \end{array} \right)  \int \frac{dk}{k^{4}}~ g_{l_1}(k)g_{l_2}(k)~,
\end{align}
where $m$ takes values in  $\{0, 2, -2\}$. The non-zero elements of $\langle a_{l_1m_1}a^*_{l_2m_2}\rangle_{2,m}$ are those with $l_1=l_2$ or $l_1 = l_2 \pm 2$ which also have comparable amplitudes.


\section{Conclusions}

In this paper, we have studied statistical anisotropies in a  model of inflation with a relic background anisotropy of the Bianchi I type. We have compared the predictions in the density perturbations in this model with another type of relic anisotropy model where the source is a vector field in the matter sector. We also considered the effect of a non-BD Gaussian state in such a model, as an illustrating example of  how the non-BD states in minimal inflation models can extend the effects of the anisotropy to shorter scales in density perturbations.

\begin{figure}[htbp]
  \centering
  \includegraphics[width=1.0\textwidth]{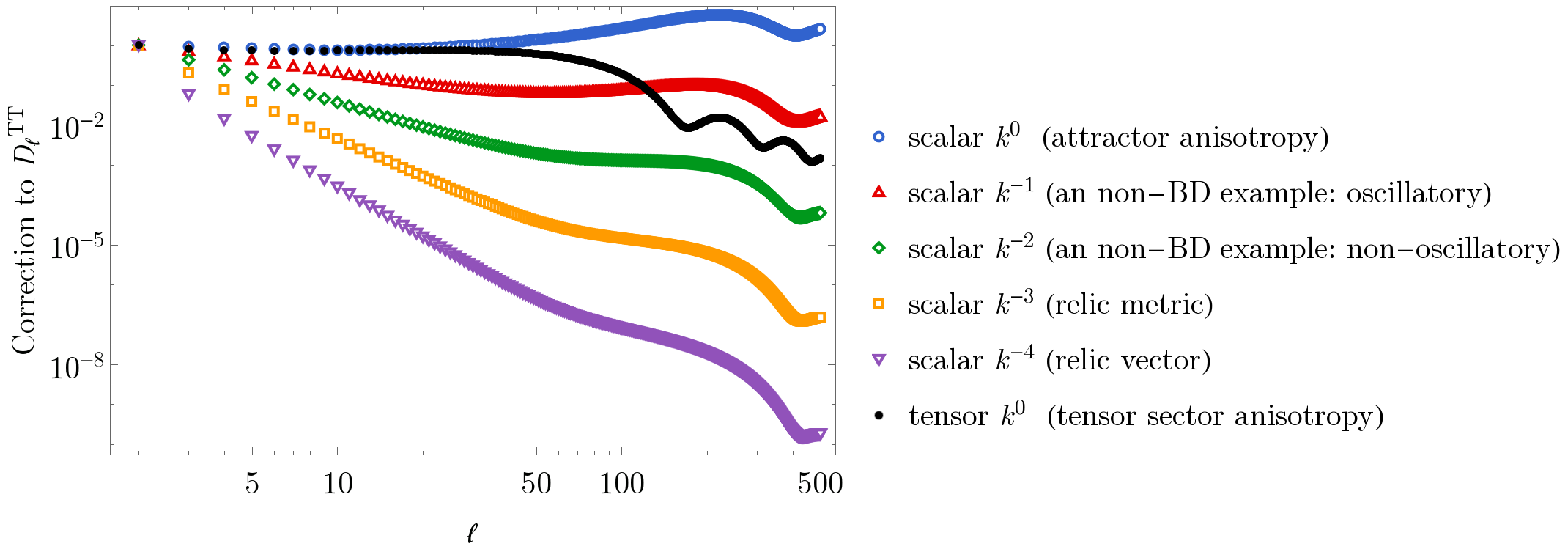}
  \caption{\label{fig:plotPower} Summary of scale-dependence of the anisotropic component in power spectrum for various models. The $k$-dependence of the primordial power spectrum, and the examples are listed in the plot legend. The correction to the CMB temperature anisotropy is plotted in the figure. For the oscillatory result in the non-BD example, only the envelop is plotted.}
\end{figure}

In Fig.~\ref{fig:plotPower} we have summarized the scale-dependence of the anisotropic power spectrum of these relic scenarios. As expected, the angular dependence of the statistical anisotropy in density perturbations are the same for all three models, because we have set up the same axial symmetric initial condition. Nonetheless, interestingly, the differences in the underlying physics of the models still lead to distinctive observable differences. In the BD vacuum case, the statistical anisotropy in the vector relic field model decays as $\sim 1/k^4$ while in the Bianchi model it decays as $\sim 1/k^3$. These behaviors are determined by the different background sources of the anisotropy. In the example of non-BD Gaussian state, the scale-dependence in both models become the same, $\sim 1/k^2$ for the non-oscillatory part, dominated by the similar initial quantum states in both cases. Such a quantum state enhances the anisotropy in much shorter scales, and becomes an interesting probe of the initial quantum state of the universe.

For comparison, in Fig.~\ref{fig:plotPower} we also listed the predictions from the models of  anisotropic inflation
based on attractor gauge field dynamics such as in \cite{Watanabe:2009ct, Emami:2010rm} in which
anisotropies are generated actively during entire period of inflation. To leading order,
the anisotropic power spectrum in these models  are given by $\delta P_\zeta \propto P_{\zeta 0} N(k)^2$ in which $N(k)$ represents the number of e-folds when the mode of interest $k$ leaves the horizon \cite{  Watanabe:2010fh, Dulaney:2010sq, Gumrukcuoglu:2010yc, Abolhasani:2013zya, Thorsrud:2012mu, Bartolo:2012sd, Shiraishi:2013vja, Nurmi:2013gpa, Fujita:2013qxa, Shiraishi:2013oqa, Lyth:2013sha}. To leading order (neglecting the logarithmic scale-dependence of $N(k)$ to $k$) the attractor anisotropic models predict nearly scale-invariant anisotropic power spectrum. As another class of scale dependence, when the anisotropies are originated from the tensor sector, the scale dependence is characterized by  the CMB transfer function from the primordial tensor mode into temperature \cite{Chen:2014eua}.

Finally we also generalized the anisotropy from the axial symmetry to arbitrary angular dependence and calculated the corresponding anisotropic power spectrum.

\section*{Acknowledgment}
XC is supported in part by a NSF grant PHY-1417421.
YW is supported by a Starting Grant of the European Research Council (ERC STG grant 279617), and the Stephen Hawking Advanced Fellowship.

\appendix{}
\section{Details of slow roll approximations}
\label{slow-roll}

In this appendix we  calculate the approximate solution of Eqs. (\ref{back-rho-eq}-\ref{anisotropy-eq}).

We start with Eq.(\ref{anisotropy-eq}). Since there is not any source of the anisotropy from the matter sector the anisotropic expansion rate decays exponentially and the non-linear term anisotropic terms in Einstein equations are  not important. Using the conformal time, $d\eta \equiv \frac{dt}{a(t)}$, the solution of Eq. (\ref{anisotropy-eq}) is
\ba
\label{sigma evo}
\sigma' = \sigma'_{0} \left(\frac{a_{0}}{a}\right)^2 \, ,
\ea
in which prime refers to the derivative respect to the conformal time and the subscript $0$ means the initial values of the corresponding quantities. Since $\sigma'$ decays very rapidly, it will not change the evolution of $\alpha$. Now by using the definition of the slow-roll parameter,
\ba
\epsilon_{H} \equiv -\frac{\ddot{\alpha}}{\dot{\alpha}^2} \, ,
\ea
we have
\ba
\left(1-\epsilon_{H} \right) = \frac{\alpha''}{\alpha'^2} \Longrightarrow \alpha' = \frac{\mathcal{H}_{0}}{1- \left(1-\epsilon_{H}\right)\mathcal{H}_{0}\left(\eta - \eta_{0} \right)} \, ,
\ea
in which we have $\mathcal{H} = a H$. Since $\mathcal{H}_{0} \eta_{0} \simeq -1$, we have
\ba
\label{scalefactor evo}
\mathcal{H} &=& \frac{\mathcal{H}_{0}}{\epsilon_{H} + \left(\epsilon_{H}-1\right)\mathcal{H}_{0}\eta} \nonumber\\
& \simeq& \left(1+\epsilon_{H}\right)\left(-\eta\right)^{-1} \, .
\ea
Integrating the above equation, we can calculate $a(\eta)$ as,
\ba
\label{a appro}
a \simeq H_{0}^{-1}\left(-\eta\right)^{-(1+ \epsilon_{H})} \, .
\ea
Now by using the above equations we can also find the evolution of $\frac{b'}{b}$. From Eq. (\ref{bian01}), written in terms of $\alpha$ and $\sigma$, we have
\ba
\frac{a'}{a} - \frac{b'}{b} = -3 \sigma' \, .
\ea
Integrating the above equation, we obtain
\ba
\label{b appro}
b &\sim & H_{0}^{-1}\left( -\eta \right)^{-(1+ \epsilon_{H})}\left( 1 + \left(\frac{\dot{\sigma_{0}}}{H_{0}}\right)\left(\mathcal{H}_{0}\eta\right)^3\right) \, .
\ea
Dropping the slow-roll parameter $\epsilon_H$ we recover Eqs. (\ref{app.a}) and (\ref{app.b}).

\section{Details of metric perturbations}
\label{metric pert}

In this appendix, we look at the perturbations of the action both from the metric and matter sectors. First  we consider the metric perturbations. Then we proceed by considering the matter sector and finally we show that, due to the hierarchy between the terms from the matter sector and the metric back-reactions, we can neglect metric perturbations and only consider the matter effects \cite{Emami:2013bk}.

\subsection{The metric perturbations}

Now we look at the perturbations of the background metric (\ref{bian0}). Since the metric components in the $x$-direction are different from the $y$ and $z$ directions, the three-dimensional rotation invariance is broken into a subset of two-dimensional rotation invariance in $y-z$ plane. Therefore, in order to classify our perturbations, one can look at the transformation properties of the physical fields under the rotation in $y-z$ plane. Therefore, we decompose all of the metric and matter perturbations into scalar and vector components with respect to the 2D rotation in the $y-z$ plane. We also note that there are no tensor perturbations in two dimensions. In order to simplify the analysis and by employing the remnant symmetry in $y-z$ plane, we put $k_{z} =0$.

With these discussions  the most general form of metric perturbations is
\ba
\label{deltag}
\delta g_{\alpha \beta} =   \left(
\begin{array}{c}
- 2 a^2 A~~~~~~~~~~~~  a^2 \partial_x \beta~~~~~~~~~~~~~~~~~~~~ a\, b \left( \partial_i B + B_i \right)
\\\\
   ~~~~~~~~~~~~~~~~~~~~~~~~      - 2 a^2 \bar \psi   ~~~~~~~~~~~~~~~~~~       a b\,  \partial_x \left( \partial_i \gamma + \Gamma_i \right)
\\\\
 ~~~~~~~~~~~~~~~~~~~~~~~~~~~~~~~~~~~~~~~~~~~~~~~~~~~~~~~~~~     b^2 \left( - 2 \psi \delta_{ij} + 2 E_{, ij} + E_{i,j} +E_{j,i} \right)
\end{array}
\right )  \, .
\ea
In this decomposition $A, \beta, B, \bar \psi, \gamma, \psi $ and $E$ are scalar perturbations while $B_i, \Gamma_i$ and $E_i$ are vector perturbations subject to transverse conditions
\ba
\label{transverse}
\partial_i E_i = \partial_ i B_i = \partial_i \Gamma_i =0 \, .
\ea
One can  choose the following gauge for the metric perturbations \cite{Emami:2013bk}:
\ba
\label{gauge0}
\psi= \bar \psi = E = E_i =0 \, .
\ea
The gauge in Eq. (\ref{gauge0}) is similar to the flat gauge in standard FRW background.

\subsection{The quadratic action }
\label{quad-action}

Here we present the quadratic action for the inflaton field and metric degrees of freedom.
Following the approach of \cite{Emami:2013bk},  the second order action for the scalar degrees of freedom  in Fourier space is
\ba
\label{S2-scalar-k}
&&S_2 = \int d \eta d^3 k \left[ b b' k_x^2 (A^* \beta + A \beta^*) + \frac{a b}{2} (\frac{a'}{a} + \frac{b'}{b}) k_y^2 (A^* B + A B^*) +  \frac{a b}{2} k_x^2 k_y^2 (\gamma^* A + \gamma A^*) - a^2 b^2 V(\phi) |A|^2
 \right. \nonumber \\ &&\left.
 - \frac{a b}{4} k_x^2 k_y^2 (\beta^* B + \beta B^*) + \frac{a' b}{2} k_x^2 k_y^2 (\gamma^* \beta + \gamma \beta^*) + \frac{a b}{4} k_x^2 k_y^2 (\gamma^* \beta' + \gamma \beta'^*)+ \frac{a^2}{4} k_x^2 k_y^2 |\beta|^2 - \frac{b^2}{4} k_x^2 k_y^2 (B^* \gamma' + B \gamma'^*)
\right. \nonumber \\ &&\left.
+ \frac{b^2}{4} (\frac{b'}{b} - \frac{a'}{a}) k_x^2 k_y^2 (\gamma^* B + \gamma B^*) + \frac{b^2}{4} k_x^2 k_y^2 |B|^2 + \frac{b^2}{4} k_x^2 k_y^2 | \gamma'|^2 - \frac{b^2}{4} (\frac{b''}{b} - \frac{a''}{a}) k_x^2 k_y^2 |\gamma|^2 -  \frac{b^2}{2} \phi' (A^* \delta \phi' + A \delta \phi'^*)
\right. \nonumber \\ &&\left.
+  \frac{b^2}{2} |\delta \phi'|^2 - \frac{b^2}{2} \phi' k_x^2 (\beta^* \delta \phi + \beta \delta \phi^*) - \frac{a b}{2} \phi' k_y^2 (B^* \delta \phi + B \delta \phi^*) - \frac{b^2}{2} k_x^2 | \delta \phi|^2 - \frac{a^2}{2} k_y^2 |\delta \phi|^2
\right. \nonumber \\ &&\left.
- \frac{a^2 b^2}{2} V_{, \phi \phi} |\delta \phi|^2  -  \frac{a^2 b^2}{2} V_{,\phi} (\delta \phi A^* + \delta \phi^* A)
\right] \, .
\ea
We have to integrate out the non-dynamical variables $\{\beta, B, A\}$ from the action Eq. (\ref{S2-scalar-k}). The analysis is simple but tedious. It turns out that it would be much easier to first integrate out $\beta$, then $B$ and finally $A$. Performing the details of integrating out analysis, the final action for the remaining dynamical field is $L= L_{\phi\phi} + L_{\gamma-\gamma} + L_{\phi \, \gamma}$ in which
\ba
\label{L_{phi-phi}}
L_{\phi \phi} &=& \frac{b^2}{2} \Big| \delta \phi ^{'}\Big|^2 - \bigg{(}\frac{b^2}{2}k^2 + \frac{a^2 b^2}{2}V_{, \phi \phi}
+\frac{b^4 k_{x}^2}{a^2 k_{y}^2} \phi^{'2} + \frac{b^6 k^4}{a^6 k_{y}^6} \frac{\phi^{'2}}{\lambda^2}\left( a^4 k_{y}^2 V(\phi) + 4 b'^2 k_{x}^2\right) + \frac{b^4 k^2}{2 a^4 k_{y}^4} \frac{\phi'}{\lambda} \times \nonumber\\
& \times & \left( 2 a^4 k_{y}^2 V_{,\phi} - 8 b b' k_{x}^2 \phi' \right)\bigg{)}\Big| \delta \phi \Big|^2
+ \bigg{(}\frac{b^4 k^2}{2 a^2 k_{y}^2} \frac{\phi'^{2}}{\lambda} \bigg{)}^{'}\Big( \delta \phi \Big)^{2} \\
\label{L_{gamma-gamma}}
L_{\gamma \gamma} & =& \frac{b^4 k_{x}^4}{a^2\lambda^2}\left( \frac{b'^2}{b^2}+ \frac{\phi'^{2}}{2}\right) \Big| \gamma ^{'}\Big|^2 - \bigg{(} \frac{b^2}{2} \frac{k_{x}^4}{\lambda}\left(k_{y}^2 - 2\frac{a'b b'}{a^3}+ 2\frac{b^{'2}}{a^2}\right)\bigg{)}^{'}\Big| \gamma \Big|^2
\ea
\ba
\label{L_{phi-gamma}}
L_{\phi \gamma} & =& \left(-\frac{b^3}{2a}k_{x}^2 \phi'-\frac{b^5 k^2 k_{x}^2}{a^5 k_{y}^4} \frac{\phi'}{\lambda^2}\left( a^4 k_{y}^2 V(\phi) + 4 b'^2 k_{x}^2\right) - \frac{b^3 k_{x}^2}{2 a^4 k_{y}^2}\frac{1}{\lambda} \left( -2k^2 \phi' a b b' + V_{,\phi}k_{y}^2 a^5 - 4 a b b' k_{x}^2\phi'\right)\right) \left(\delta \phi^{*}\gamma ^{'} + c.c.\right)+ \nonumber\\
& + &\bigg{(}\frac{b^3 k_{x}^2}{2a}\phi'\left(\frac{a^{'}}{a}-\frac{b^{'}}{b}\right)
-\frac{b^3 k_{x}^2}{2 a^4 k_{y}^2}\frac{1}{\lambda}\left( -a^3 k^2 k_{y}^2 \phi' + 2 a' b b'k^2 \phi'- 2 a b'^2 k^2 \phi'\right)\bigg{)}\bigg{(}\delta \phi^{*}\gamma + c.c.\bigg{)} + \nonumber\\
&-& \frac{b^3 k_{x}^2}{2 a}\frac{\phi'}{\lambda}\bigg{(}\delta \phi^{'*}\gamma^{'} + c.c.\bigg{)}
\ea
Note that  we have defined $k$ as, $k^2 \equiv k_{x}^2 + \frac{a^2}{b^2} k_{y}^2$. (Note that there is a clash of notation here, this definition of $k$ is different from those in the main text defined as Eq.~(\ref{k_definition}).)
In addition, $\lambda$ has been defined as
\ba
\lambda \equiv \frac{a'}{a} + \frac{b'}{b} + 2 \frac{b^2}{a^2}\frac{k_{x}^2}{k_{y}^2}\frac{b'}{b} \, .
\ea

\subsection{Leading Correction}
To see the leading corrections in the action let us take a look at Eq. (\ref{L_{phi-gamma}}). As we can see all of the terms are proportional to $\phi'$ which means that they are all slow-roll suppressed. These terms are due to the metric perturbations since in the original action Eq.~(\ref{S2-scalar-k}) there is not any mixing between $\phi$ and $\gamma$.  The situation for Eq.~(\ref{L_{phi-phi}}) is the same, terms that are not directly from the matter sector are proportional to $\phi'$ or $V_{,\phi\phi}$ and so are slow-roll suppressed.  Therefore, we conclude that the metric perturbations in quadratic action are sub-leading compared to the contributions from the matter sector
fluctuations.


\section{Detail analysis of $\psi$}
\label{psi solution}

Here we write down the equation of motion for $\psi$, which is defined by Eq. (\ref{psi u}), and try to solve it perturbatively.
\ba
\label{psi eq}
\psi''_{k} + \psi'^2_{k} - \frac{2}{\eta}\left( 1 - 3\left(\frac{\dot{\sigma_{0}}}{H_{0}}\right) (\mathcal{H}_{0}\eta)^3 \right)\psi'_{k} + \left[ k^2 - 2\left(\frac{\dot{\sigma_{0}}}{H_{0}}\right)(\mathcal{H}_{0}\eta)^3 \left(k_{y}^2 + k_{z}^2\right) \right] \psi_k =0 \, .
\ea
Expanding $\psi$ as
\ba
\psi_{\mathbf{k}}(\eta) = \psi_{\mathbf{k}(0)}(\eta)+ \psi_{\mathbf{k}(1)}(\eta)+ ...
\ea
the first order equation of motion for $\psi$ is
\ba
\psi''_{k(1)} + 2 \psi'_{k(0)}\psi'_{k(1)} - \frac{2}{\eta}\psi'_{k(1)} + 2 \left(\frac{\dot{\sigma_{0}}}{H_{0}}\right)(\mathcal{H}_{0}\eta)^3 \left( \frac{3}{\eta}\psi'_{k(0)} - \left(k_{y}^2 + k_{z}^2\right) \right) = 0
\ea
We can solve this equation and use the normalization condition to fix the constant of integration. The final result is
\ba
\psi_{\mathbf{k}(1)}(\eta) &=& \frac{\mathcal{H}_{0}^2 \sigma_{0}'}{1+ik\eta}\sum_{n=0}^{5} {\beta_{n}}\eta^{n}
\ea
Where ${\beta_{+n}}$ are given by
\ba
{\beta_{0}} &=& \frac{3i}{8k^3}\left(1 + 3 \cos^2{\Theta}\right) \\
{\beta_{1}} &=& -\frac{3}{8k^2}\left(1 + 3 \cos^2{\Theta}\right) \\
{\beta_{2}} &=& 0 \\
{\beta_{m}} &=& \alpha_{m} ~~,~~ (m=3,4,5) \, .
\ea


\section{Summation rules of spherical harmonics}
\label{sec:summ-rules-spher}

In the following, we present  a general expression for the $l$-dependence of the diagonal part of $ C_{l}$ due to a general anisotropic model,
\begin{align} \label{summation}
\frac{1}{2l+1}  \sum_m \int d\Omega_k ~ Y_{lm}(\mathbf{k}) Y^*_{lm}(\mathbf{k})Y_{LM}(\mathbf{k}) = \sqrt{\frac{2L+1}{4\pi}}\left( \begin{array}{ccc}
      l & l & L\\
      0 & 0 & 0
    \end{array} \right)\sum_m (-1)^{m}
  \left( \begin{array}{ccc}
      l & l & L\\
      -m & m & M
    \end{array} \right) ~.
\end{align}
It is worth to simplify Eq. (\ref{summation}). We first note that, due to the conservation of angular momentum, we have $M=0$. In addition, we can use the following identity,
\begin{align}
\sum_m (-1)^{m}
  \left( \begin{array}{ccc}
      l & l & L\\
      -m & m & 0
    \end{array} \right) = (-1)^{l}\sqrt{2l+1}\delta_{L0} ~.
\end{align}
So
\begin{align}
\frac{1}{2l+1}  \sum_m \int d\Omega_k ~ Y_{lm}(\mathbf{k}) Y^*_{lm}(\mathbf{k})Y_{LM}(\mathbf{k}) = & \sqrt{\frac{2L+1}{4\pi}} \delta_{L0}(-1)^{l}\sqrt{2l+1}\left( \begin{array}{ccc}
      l & l & L\\
      0 & 0 & 0
    \end{array} \right)\nonumber\\
    =& \sqrt{\frac{1}{4\pi}}(-1)^{l}\sqrt{2l+1}\left( \begin{array}{ccc}
      l & l & 0\\
      0 & 0 & 0
    \end{array} \right) ~.
\end{align}
Finally by using the following identity,
\begin{align}
\left( \begin{array}{ccc}
      l & l & 0\\
      0 & 0 & 0
    \end{array} \right) = (-1)^{l}\frac{1}{\sqrt{2l+1}} ~,
\end{align}
we get
\begin{align} \label{summation1}
\frac{1}{2l+1}  \sum_m \int d\Omega_k ~ Y_{lm}(\mathbf{k}) Y^*_{lm}(\mathbf{k})Y_{LM}(\mathbf{k}) = \sqrt{\frac{1}{4\pi}} \delta_{L0}\delta_{M0} ~.
\end{align}



\section*{References}
\begin{thebibliography}{}



\bibitem{Bennett:2010jb}
  C.~L.~Bennett {\it et al.},
  Astrophys.\ J.\ Suppl.\  {\bf 192}, 17 (2011)
  [arXiv:1001.4758 [astro-ph.CO]].


\bibitem{Ade:2013nlj}
  P.~A.~R.~Ade {\it et al.}  [Planck Collaboration],
  arXiv:1303.5083 [astro-ph.CO].


\bibitem{Tegmark:2003ve}
  M.~Tegmark, A.~de Oliveira-Costa and A.~Hamilton,
  Phys.\ Rev.\ D {\bf 68}, 123523 (2003)
  [astro-ph/0302496].

\bibitem{de OliveiraCosta:2003pu}
  A.~de Oliveira-Costa, M.~Tegmark, M.~Zaldarriaga and A.~Hamilton,
  Phys.\ Rev.\ D {\bf 69}, 063516 (2004)
  [astro-ph/0307282].

\bibitem{Schwarz:2004gk}
  D.~J.~Schwarz, G.~D.~Starkman, D.~Huterer and C.~J.~Copi,
  Phys.\ Rev.\ Lett.\  {\bf 93}, 221301 (2004)
  [astro-ph/0403353].

\bibitem{Land:2005ad}
  K.~Land and J.~Magueijo,
  Phys.\ Rev.\ Lett.\  {\bf 95}, 071301 (2005)
  [astro-ph/0502237].

\bibitem{Copi:2006tu}
  C.~Copi, D.~Huterer, D.~Schwarz and G.~Starkman,
  Phys.\ Rev.\ D {\bf 75}, 023507 (2007)
  [astro-ph/0605135].



\bibitem{Gumrukcuoglu:2007bx}
  A.~E.~Gumrukcuoglu, C.~R.~Contaldi and M.~Peloso,
  JCAP {\bf 0711}, 005 (2007)
  [arXiv:0707.4179 [astro-ph]].

\bibitem{Pereira:2007yy}
  T.~S.~Pereira, C.~Pitrou and J.~-P.~Uzan,
  JCAP {\bf 0709}, 006 (2007)
  [arXiv:0707.0736 [astro-ph]].


\bibitem{Chen:2013tna}
  X.~Chen and Y.~Wang,
  JCAP {\bf 1407} 004 (2014)
  [arXiv:1306.0609 [hep-th]].


\bibitem{Chen:2013eaa}
  X.~Chen and Y.~Wang,
  arXiv:1305.4794 [astro-ph.CO].


\bibitem{Watanabe:2009ct}
  M.~a.~Watanabe, S.~Kanno and J.~Soda,
  Phys.\ Rev.\ Lett.\  {\bf 102}, 191302 (2009)
  [arXiv:0902.2833 [hep-th]].

\bibitem{Emami:2010rm}
  R.~Emami, H.~Firouzjahi, S.~M.~Sadegh Movahed, M.~Zarei,
  JCAP {\bf 1102 } (2011)  005.
  [arXiv:1010.5495 [astro-ph.CO]].


\bibitem{Ohashi:2013qba}
  J.~Ohashi, J.~Soda and S.~Tsujikawa,
  JCAP {\bf 1312}, 009 (2013)
  [arXiv:1308.4488 [astro-ph.CO], arXiv:1308.4488].

\bibitem{Thorsrud:2013kya}
  M.~Thorsrud, F.~R.~Urban and D.~F.~Mota,
  JCAP {\bf 1404}, 010 (2014)
  [arXiv:1312.7491 [astro-ph.CO]].

\bibitem{Chen:2014eua}
  X.~Chen, R.~Emami, H.~Firouzjahi and Y.~Wang,
  arXiv:1404.4083 [astro-ph.CO].

\bibitem{Li:2009sp}
  M.~Li and Y.~Wang,
  JCAP {\bf 0907}, 033 (2009)
  [arXiv:0903.2123 [hep-th]].

\bibitem{Afshordi:2010wn}
  N.~Afshordi, A.~Slosar and Y.~Wang,
  JCAP {\bf 1101}, 019 (2011)
  [arXiv:1006.5021 [astro-ph.CO]].

\bibitem{Wang:2013zz}
  Y.~Wang,
  JCAP {\bf 1310}, 006 (2013)
  [arXiv:1304.0599 [astro-ph.CO]].

\bibitem{Endlich:2012pz}
  S.~Endlich, A.~Nicolis and J.~Wang,
  JCAP {\bf 1310}, 011 (2013)
  [arXiv:1210.0569 [hep-th]].

\bibitem{Bartolo:2013msa}
  N.~Bartolo, S.~Matarrese, M.~Peloso and A.~Ricciardone,
  JCAP {\bf 1308}, 022 (2013)
  [arXiv:1306.4160 [astro-ph.CO]].

\bibitem{Akhshik:2014zz}
 M.~Akhshik, R.~Emami, H.~Firouzjahi, Y.~Wang,
  [arXiv:1405.4179 [astro-ph.CO]].

\bibitem{Dey:2013tfa}
  A.~Dey, E.~D.~Kovetz and S.~Paban,
  JCAP {\bf 1406}, 025 (2014)
  [arXiv:1311.5606 [hep-th]].

\bibitem{Dey:2011mj}
  A.~Dey and S.~Paban,
  JCAP {\bf 1204}, 039 (2012)
  [arXiv:1106.5840 [hep-th]].

\bibitem{Dey:2012qp}
  A.~Dey, E.~Kovetz and S.~Paban,
  JCAP {\bf 1210}, 055 (2012)
  [arXiv:1205.2758 [astro-ph.CO]].

\bibitem{Emami:2014tpa}
  R.~Emami, H.~Firouzjahi and M.~Zarei,
  Phys.\ Rev.\ D {\bf 90}, 023504 (2014)
  [arXiv:1401.4406 [hep-th]].

\bibitem{Emami:2013bk}
  R.~Emami and H.~Firouzjahi,
  JCAP {\bf 1310}, 041 (2013)
  [arXiv:1301.1219 [hep-th]].

\bibitem{Polarski:1995jg}
  D.~Polarski and A.~A.~Starobinsky,
  Class.\ Quant.\ Grav.\  {\bf 13}, 377 (1996)
  [gr-qc/9504030].


\bibitem{Barrow:1997sy}
  J.~D.~Barrow,
  Phys.\ Rev.\ D {\bf 55}, 7451 (1997)
  [gr-qc/9701038].

\bibitem{Barrow:1998ih}
  J.~D.~Barrow and R.~Maartens,
  Phys.\ Rev.\ D {\bf 59}, 043502 (1999)
  [astro-ph/9808268].


\bibitem{Brandenberger:2000wr}
  R.~H.~Brandenberger and J.~Martin,
  Mod.\ Phys.\ Lett.\ A {\bf 16}, 999 (2001)
  [astro-ph/0005432].

\bibitem{Easther:2001fi}
  R.~Easther, B.~R.~Greene, W.~H.~Kinney and G.~Shiu,
  Phys.\ Rev.\ D {\bf 64}, 103502 (2001)
  [hep-th/0104102].

\bibitem{Chen:2006nt}
  X.~Chen, M.~-x.~Huang, S.~Kachru and G.~Shiu,
  JCAP {\bf 0701}, 002 (2007).

\bibitem{Holman:2007na}
  R.~Holman and A.~J.~Tolley,
  JCAP {\bf 0805}, 001 (2008).

\bibitem{Meerburg:2009ys}
  P.~D.~Meerburg, J.~P.~van der Schaar and P.~S.~Corasaniti,
  JCAP {\bf 0905}, 018 (2009).

\bibitem{Chen:2009bc}
  X.~Chen, B.~Hu, M.~-x.~Huang, G.~Shiu and Y.~Wang,
  JCAP {\bf 0908}, 008 (2009)
  [arXiv:0905.3494 [astro-ph.CO]].


\bibitem{Agullo:2010ws}
  I.~Agullo and L.~Parker,
  Phys.\ Rev.\ D {\bf 83}, 063526 (2011).

\bibitem{Ganc:2011dy}
  J.~Ganc,
  Phys.\ Rev.\ D {\bf 84}, 063514 (2011).


\bibitem{Chialva:2011hc}
  D.~Chialva,
  JCAP {\bf 1210}, 037 (2012).

\bibitem{Berezhiani:2014kga}
  L.~Berezhiani and J.~Khoury,
  arXiv:1406.2689 [hep-th].



\bibitem{Pullen:2007tu}
  A.~R.~Pullen and M.~Kamionkowski,
  Phys.\ Rev.\ D {\bf 76}, 103529 (2007)
  [arXiv:0709.1144 [astro-ph]].

\bibitem{Ma:2011ii}
  Y.~-Z.~Ma, G.~Efstathiou and A.~Challinor,
  Phys.\ Rev.\ D {\bf 83}, 083005 (2011)
  [arXiv:1102.4961 [astro-ph.CO]].

\bibitem{Kim:2013gka}
  J.~Kim and E.~Komatsu,
  Phys.\ Rev.\ D {\bf 88}, 101301 (2013)
  [arXiv:1310.1605 [astro-ph.CO]].

\bibitem{Watanabe:2010bu}
  M.~-a.~Watanabe, S.~Kanno and J.~Soda,
  Mon.\ Not.\ Roy.\ Astron.\ Soc.\  {\bf 412}, L83 (2011)
  [arXiv:1011.3604 [astro-ph.CO]].

\bibitem{Book:2011na}
  L.~G.~Book, M.~Kamionkowski and T.~Souradeep,
  Phys.\ Rev.\ D {\bf 85}, 023010 (2012)
  [arXiv:1109.2910 [astro-ph.CO]].



\bibitem{Watanabe:2010fh}
  M.~a.~Watanabe, S.~Kanno and J.~Soda,
  Prog.\ Theor.\ Phys.\  {\bf 123}, 1041 (2010)
  [arXiv:1003.0056 [astro-ph.CO]].

\bibitem{Dulaney:2010sq}
  T.~R.~Dulaney, M.~I.~Gresham,
  Phys.\ Rev.\  {\bf D81}, 103532 (2010).
  [arXiv:1001.2301 [astro-ph.CO]].


\bibitem{Gumrukcuoglu:2010yc}
  A.~E.~Gumrukcuoglu, B.~Himmetoglu, M.~Peloso,
  Phys.\ Rev.\  {\bf D81}, 063528 (2010).
  [arXiv:1001.4088 [astro-ph.CO]].


\bibitem{Abolhasani:2013zya}
  A.~A.~Abolhasani, R.~Emami, J.~T.~Firouzjaee and H.~Firouzjahi,
  JCAP {\bf 1308}, 016 (2013)
  [arXiv:1302.6986 [astro-ph.CO]].


\bibitem{Thorsrud:2012mu}
  M.~Thorsrud, D.~F.~Mota and S.~Hervik,
  JHEP {\bf 1210}, 066 (2012)
  [arXiv:1205.6261 [hep-th]].


\bibitem{Bartolo:2012sd}
  N.~Bartolo, S.~Matarrese, M.~Peloso and A.~Ricciardone,
  Phys.\ Rev.\ D {\bf 87}, 023504 (2013)
  [arXiv:1210.3257 [astro-ph.CO]].


\bibitem{Shiraishi:2013vja}
  M.~Shiraishi, E.~Komatsu, M.~Peloso and N.~Barnaby,
  JCAP {\bf 1305}, 002 (2013)
  [arXiv:1302.3056 [astro-ph.CO]].

\bibitem{Nurmi:2013gpa}
  S.~Nurmi and M.~S.~Sloth,
  arXiv:1312.4946 [astro-ph.CO].


\bibitem{Fujita:2013qxa}
  T.~Fujita and S.~Yokoyama,
  JCAP {\bf 1309}, 009 (2013)
  [arXiv:1306.2992 [astro-ph.CO]].

\bibitem{Shiraishi:2013oqa}
  M.~Shiraishi, E.~Komatsu and M.~Peloso,
  arXiv:1312.5221 [astro-ph.CO].


\bibitem{Lyth:2013sha}
  D.~H.~Lyth and M.~Karciauskas,
  JCAP {\bf 1305}, 011 (2013)
  [arXiv:1302.7304 [astro-ph.CO]].




\end{thebibliography}
\end{document}